\pdfoutput=1
\RequirePackage{iftex}
\RequirePackage{ifpdf}
\ifpdf\else\ifXeTeX\else
  \errmessage{Wrong compiler: use pdfLaTeX (Overleaf: Menu > Compiler > pdfLaTeX). DVI-route LaTeX breaks figures and URLs}%
\fi\fi
\documentclass[sigplan,twocolumn,nonacm]{acmart}
\renewcommand\footnotetextcopyrightpermission[1]{}
\usepackage{booktabs}
\usepackage{listings}
\IfFileExists{xurl.sty}{\usepackage{xurl}}{%
  \expandafter\def\expandafter\UrlBreaks\expandafter{\UrlBreaks
    \do\a\do\b\do\c\do\d\do\e\do\f\do\g\do\h\do\i\do\j
    \do\k\do\l\do\m\do\n\do\o\do\p\do\q\do\r\do\s\do\t
    \do\u\do\v\do\w\do\x\do\y\do\z
    \do\A\do\B\do\C\do\D\do\E\do\F\do\G\do\H\do\I\do\J
    \do\K\do\L\do\M\do\N\do\O\do\P\do\Q\do\R\do\S\do\T
    \do\U\do\V\do\W\do\X\do\Y\do\Z
    \do\0\do\1\do\2\do\3\do\4\do\5\do\6\do\7\do\8\do\9}%
}
\begin{document}

\title{TEE Anchor: Cross-TEE Organizational Endorsement for Mitigating TEE Physical Attacks}

%% Anonymized for double-blind review. Restore for camera-ready:
% \author{Ao Sakurai}
% \email{ao.sakurai@acompany-ac.com}
% \affiliation{%
%   \institution{Confidential Computing Lab, Acompany Co., Ltd.}
%   \city{Nagoya}
%   \country{Japan}}
\author{Ao Sakurai}
\email{ao.sakurai@acompany-ac.com}
\affiliation{%
  \institution{Confidential Computing Lab, Acompany Co., Ltd.}
  \city{Nagoya}
  \country{Japan}}

\begin{abstract}
In 2025, practical physical-access attacks against TEEs, such as TEE.fail and Battering RAM, were disclosed, posing a serious threat to current TEEs. The more strictly the target machine is guarded, the harder such attacks are to mount. \emph{Residing under trusted management} has therefore emerged as a new security requirement. Yet existing attestation cannot prove a machine's organizational affiliation. This leaves room for an attacker to pass off a physically accessible machine under their control as one operated in a legitimate environment. We propose TEE Anchor, a mechanism that mitigates physical attacks by proving which organization manages a machine. TEE Anchor is a lightweight, X.509-based design that needs no additional root of trust such as a (v)TPM. The organization issues a certificate over the Chip ID, a unique per-CPU identifier, forming its own PKI. A Verifier then proves affiliation by matching the Chip ID in the attestation evidence against that certificate. This outperforms (v)TPM-based prior work in operational cost and deployability, applies across major TEEs without vendor lock-in, and lets any organization assert affiliation independently of TEE vendors. We implemented a prototype and confirmed that all operations, including provisioning and verification, complete within a few milliseconds.
\end{abstract}




\maketitle

\section{Introduction}
\subsection{Trusted Execution Environments}
Trusted Execution Environments (TEEs) protect data in use by isolating designated regions of main memory through CPU access control and, in many designs, additionally encrypting them with an in-CPU memory encryption engine~\cite{ccc-wp}. This shields code and data from untrusted parties, including privileged software such as a cloud provider's OS, regardless of privilege level. The current generation of TEEs comprises Scalable-SGX, the next-generation SGX for server CPUs, and Confidential VM (CVM) designs such as Intel TDX, AMD SEV-SNP, and Arm CCA, which place an entire virtual machine under protection.

\subsection{Remote Attestation}
Remote Attestation (RA) lets a remote user verify that a TEE is running correctly on a genuine machine manufactured by the CPU vendor with the expected behavior definition and security level, establishing the authenticity and integrity of the TEE instance. Following standard terminology~\cite{rats}, the remote user is called the Relying Party (RP), the TEE under verification is called the Attester, and the verifying entity is called the Verifier. The RP may also act as the Verifier. In RA, metadata describing the state of the TEE and the machine is kept within the Trusted Computing Base (TCB), the portion of the system that must be assumed trustworthy for the TEE to operate correctly. This metadata is signed with an Attestation Key (AK) derived from a secret held in the machine's Root of Trust, and the resulting signed structure is the Attestation Report (AR). Since the AK or its derivation secret is known to the TEE vendor, the vendor endorses it via a certificate chain anchored in its own root CA, which lets a Verifier confirm that an AR originates from a genuine TEE machine.

\subsection{Physical Attacks on TEEs}
In 2025, TEE.fail~\cite{teefail} and Battering RAM~\cite{batteringram} were disclosed as practical physical-access attacks against TEEs. Both attacks insert an attack device called an interposer between a machine's DIMMs (main memory) and the DIMM slots. TEE.fail observes main-memory traffic through the interposer to capture ciphertexts of TEE memory and then mounts a ciphertext side-channel attack to extract secrets. Battering RAM uses the interposer to construct aliases from multiple memory addresses to a single physical location in DRAM, and exploits this aliasing to extract secrets.

A major root cause common to both attacks is that current TEEs employ deterministic memory encryption using algorithms such as AES-XTS or AES-XEX. These algorithms take the physical address as the tweak, so identical data placed at the same physical address yields the same ciphertext as long as the encryption key is unchanged. Both attacks abuse this deterministic behavior. Fixing this requires hardware-level modification of the in-CPU memory encryption engine. Moreover, the consumer-oriented Client-SGX, which resisted this attack class thanks to a tweak that varied on every write, has already been discontinued partly for performance reasons~\cite{batteringram}, so waiting for robust hardware-level fixes is not realistic as of 2026.

Notably, these physical attacks require direct installation of an interposer on the machine. Conversely, if a malicious actor cannot physically access the machine, such attacks become impossible to mount. This means that machines operated in tightly guarded locations, such as data centers managed by cloud providers, enjoy higher resistance to physical attacks. Therefore, until fundamental countermeasures are deployed, verifying \emph{where a machine resides}, that is, which organization manages it, serves as an indirect but effective defense. Existing RA, however, cannot verify this organizational affiliation.

\subsection{Contributions}
To address this problem, we propose TEE Anchor, a mechanism for affiliation assurance that consists of two phases. In the \emph{provisioning} phase, the organization managing a machine registers that machine with itself, and in the \emph{verification} phase, a Verifier checks that the Attester machine truly belongs to that organization. The current major TEEs, namely SGX, TDX, SEV-SNP, and CCA, each expose a CPU-unique identifier, which we collectively call the Chip ID.   

In provisioning, the organization first validates the tamper-proof evidence that carries the Chip ID, namely the AR or the TEE-vendor certificates used to verify it, and extracts the Chip ID from it. The organization then signs the Chip ID with a key anchored in its own root CA private key, forming an organization-specific PKI whose certificate chain is published to Verifiers. Certificates can also be revoked upon compromise or decommissioning of machines. Affiliation could instead be endorsed via the AK public key, but AKs rotate on every TCB update, which would severely harm maintainability. We therefore adopt the Chip ID. 

In verification, the Verifier performs conventional RA and then validates the organizational chain, comparing the Chip ID in the organizational leaf certificate with the one in the verified AR or vendor certificates. If the two match, the Verifier concludes that the machine belongs to the organization. Because certificates are bound to the CPU-unique Chip ID, presenting an AR relayed from a different machine does not succeed.

TEE Anchor is a lightweight X.509-based design that depends on no additional root of trust such as a (v)TPM. Since provisioning is carried out by the organization itself, any organization independent of TEE vendors, such as a cloud provider, a financial institution, or a Web3 community, can assert affiliation without vendor lock-in.

Our contributions are summarized as follows:
\begin{itemize}
    \item We present TEE Anchor, which enables any trustworthy organization to build an X.509-based PKI over the Chip IDs of its TEE machines. By verifying this PKI in addition to conventional RA, a Verifier can confirm that a machine resides in a secure environment where physical attacks are infeasible.
    \item TEE Anchor is non-invasive to existing RA and needs no delegation to third parties, because its verification is layered on top of conventional RA. It applies to any TEE whose Chip ID can be extracted securely, and therefore needs no costly stacks such as TPMs. It thus excels in both deployability and performance, without vendor lock-in.
    \item We implemented a provisioning and verification tool covering Scalable-SGX, TDX, SEV-SNP, and CCA, experimentally confirmed very small overhead, and showed that it resolves problems faced by prior work.
    \item We discuss how TEE Anchor's affiliation guarantee can also be used to mitigate the known diversion and session-binding relay attacks on Attested TLS.
\end{itemize}

\section{Threat Model}\label{sec:threat}
\subsection{Attacker Model}
We assume an attacker who possesses a TEE machine they can physically access and who can mount physical attacks such as TEE.fail or Battering RAM against that machine. Such attacks expose TEE memory in plaintext. On TEEs whose AK is held in main memory, such as SGX/TDX, the attacker can therefore extract the AK itself and forge ARs. On the other hand, the attacker cannot physically access machines managed by the target trustworthy organization. The attacker also has the standard ability to observe public information on the network, such as published PKI material and intercepted traffic.

The attacker's goal is to make a Verifier believe that a workload is running on a TEE managed by the target organization while actually executing it on a machine the attacker can physically access, disguising the results as coming from an organization-managed machine. By then mounting a physical attack, the attacker defeats the TEE's confidentiality and integrity guarantees, e.g., stealing confidential data or tampering with results.

\subsection{Proxy/Relay Attacks}\label{sec:proxyrelay}
The attacker attempts to bypass legitimate verification as follows:
\begin{enumerate}
    \item Launch a TEE instance on the physically attackable machine at hand and obtain a genuine AR.
    \item Separately obtain the organizational certificate chain published by the target organization.
    \item Present the AR and the organizational certificate chain to the Verifier.
    \item The Verifier fails to detect that the two originate from different physical machines and the verification passes.
\end{enumerate}
We call this class of attacks Proxy/Relay attacks.

\subsection{Out-of-Scope Threats}
First, we do not consider attackers who can physically access organization-managed machines, which could happen through organizational mismanagement or supply-chain hardware compromise. The organization bears responsibility for its physical security and procurement channels. It may also reset, at deployment time, any platform information that could later be abused to forge ARs, using facilities such as SGX Factory Reset.

\subsection{Verification Goals}\label{sec:goals}
Under this threat model, a verification procedure must satisfy three requirements. First, the evidence of affiliation must be bound to the same physical machine as the AR, so that the Proxy/Relay attacks of Section~\ref{sec:proxyrelay} fail. Second, the Verifier must be able to reach its decision without trusting any party other than the TEE vendor and the endorsing organization, so that affiliation assurance does not widen the set of trusted parties. Third, the organization must be able to withdraw its endorsement of a machine without waiting for the TEE vendor. The latter two keep the trust and the operational cost that affiliation assurance adds to conventional RA to a minimum.

\section{Design}\label{sec:design}
TEE Anchor requires no modification to TEE hardware, firmware, or the conventional RA. It adds only an organizational PKI on the provisioning side and one extra verification step on the Verifier side, both of which we describe below.

\subsection{Provisioning Phase}
\begin{figure}[t]
    \centering
    \includegraphics[width=\linewidth]{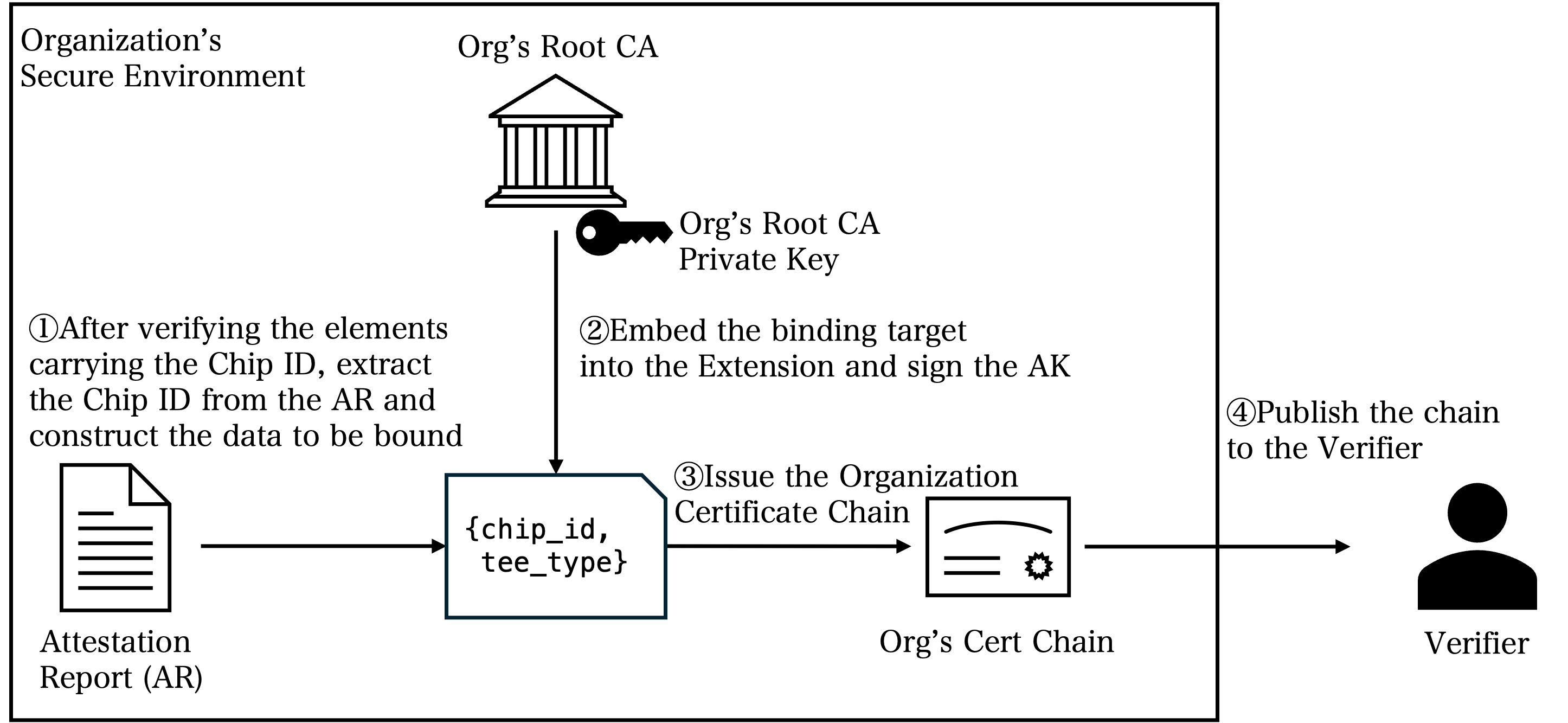}
    \caption{Overview of the provisioning phase in TEE Anchor.}
    \label{fig:provision}
\end{figure}

Figure~\ref{fig:provision} shows an overview of TEE Anchor's provisioning phase. As the first step, the organization's administrator extracts the Chip ID from a tamper-proof, TEE-related data structure of the machine. As Chip IDs, the PPID (Platform Provisioning ID) can be used for SGX/TDX, \texttt{CHIP\_ID} for SEV-SNP, and \texttt{cca-platform-instance-id} for CCA. For SGX/TDX, the Provisioning Certification Key (PCK) is a kind of AK, and the TEE vendor issues a PCK Certificate for it. The PPID is stored in an X.509 extension of that certificate, which is normally attached to the Quote, the AR format of SGX/TDX. For SEV-SNP, \texttt{CHIP\_ID} is a member of the AR itself, and for CCA, \texttt{cca-platform-instance-id} is contained in the Platform Token, a component of the AR.  

The administrator first validates the enclosing TEE vendor certificate or AR against the TEE vendor's root CA certificate, and then extracts the Chip ID. For SGX/TDX, since our threat model assumes the Quote is obtained legitimately in the organization's secure environment, a mismatch between the Quote and its attached PCK Certificate cannot arise. We therefore check only the PCK Certificate at provisioning time and leave Quote verification to conventional RA.

Next, the administrator stores the extracted Chip ID and the TEE type in X.509 extensions of a leaf certificate. The TEE type, such as SGX or CCA, is specified by the administrator through TEE Anchor's command interface. Since X.509 requires a Subject Key, the leaf certificate embeds the TEE's AK public key as a placeholder. Because this key is not used in the verification decision, a dummy key would also suffice.   

The organization then signs the leaf certificate with a key anchored in its root CA private key, and issues an organizational certificate chain that consists of the leaf certificate and one or more superior certificates. The chain is published to Verifiers by any suitable means. Efficient distribution of the chains is out of scope for this paper.

\subsubsection{Revocation}
When a provisioned machine is decommissioned or sold, it leaves the organization's management, so the corresponding certificate must be revoked. Moreover, if a machine is compromised by any attack, conventional RA-side revocation first requires reporting to the TEE vendor, and it may take time for the vendor to reflect the revocation. It is faster for the organization to revoke the certificate on its own. TEE Anchor therefore includes the ability to revoke certificates for TEE machines. A Verifier can receive the organizational certificate revocation list (CRL) from the organization and use it for revocation checks.

\subsection{Verification Phase}
\begin{figure}[t]
    \centering
    \includegraphics[width=\linewidth]{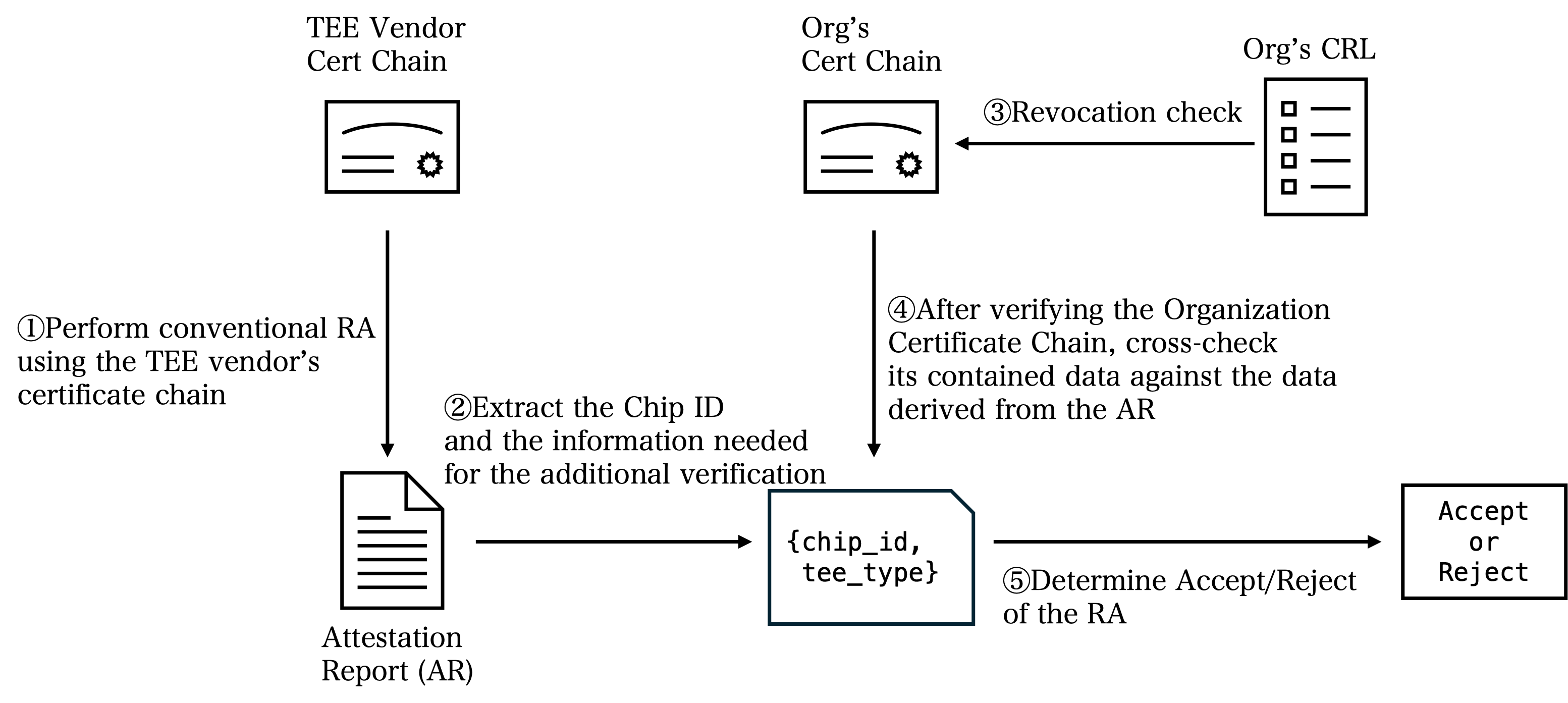}
    \caption{Overview of the verification phase in TEE Anchor.}
    \label{fig:verify}
\end{figure}

%% NOTE: figure* floats to the top of the NEXT page after its point of
%% encounter in two-column mode; placed here (Sec. 5 source, typeset on p.4)
%% so that it lands on the top of p.5.
\begin{figure*}[t!]
    \centering
    \includegraphics[width=\textwidth]{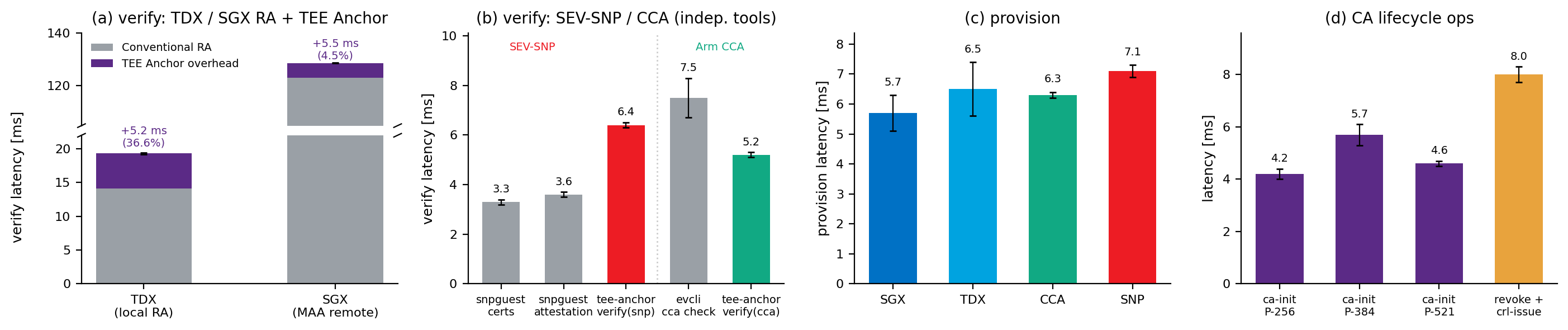}
    \caption{Every TEE Anchor operation completes within roughly 5--8\,ms, so affiliation assurance adds only a few milliseconds to attestation. (a)~overhead of \texttt{verify} over conventional RA on TDX/SGX; (b)~comparison of conventional RA and \texttt{verify} on SEV-SNP/CCA; (c)~\texttt{provision}; (d)~CA lifecycle operations (\texttt{revoke} + \texttt{crl-issue}). Error bars show standard deviation.}
    \label{fig:bench}
\end{figure*}

Figure~\ref{fig:verify} shows an overview of the verification phase. The Verifier first performs conventional RA on the Attester's TEE using the AR and the TEE vendor certificate chain. If conventional RA already warrants rejection, the procedure terminates there. If the conventional RA result is acceptable, the Verifier next extracts the Chip ID from the verified AR or from the TEE vendor certificates used in its verification. The Verifier then validates the organizational certificate chain against the organizational root CA certificate. On success, the Verifier extracts the Chip ID and the TEE type from the extensions of the organizational leaf certificate. The Chip ID derived from the AR is then compared with the one derived from the organizational certificate. If the two match, the Verifier gains confidence that the Attester's machine truly belongs to that organization. It also compares the TEE types as a sanity check that the designated TEE type is being verified. Because the organizational certificate is bound to the Chip ID, an AR relayed from another machine cannot match it, which defeats the Proxy/Relay attacks of Section~\ref{sec:proxyrelay}. This holds even where the attacker can forge ARs with a stolen AK, since the TEE vendor's endorsement binds that AK to the Chip ID of the machine it came from.

Optionally, the Verifier can also supply the organizational CRL during this verification, and thereby reject machines that the organization has revoked.

\section{Implementation}\label{sec:impl}
We implemented a prototype of the proposed scheme as a standalone command-line tool, \texttt{tee-anchor}. It uses OpenSSL for generating and signing organizational certificates, handling X.509 extensions, and issuing CRLs, and provides the five subcommands shown in Table~\ref{tab:subcommands}. Here, \texttt{provision} and \texttt{verify} correspond to the two phases of Section~\ref{sec:design}, \texttt{revoke} and \texttt{crl-issue} implement the revocation capability described there, and \texttt{ca-init} sets up the organizational CA. It supports Intel Scalable-SGX and TDX, AMD SEV-SNP, and Arm CCA, internally absorbing the differences among the vendor-specific structures carrying the Chip ID (PCK Certificate, Attestation Report, and Platform Token).
\begin{table}[t]
  \centering
  \caption{Subcommands provided by \texttt{tee-anchor}.}
  \label{tab:subcommands}
  \begin{tabular}{@{}l p{160pt}@{}}
    \toprule
    Subcommand & Function \\
    \midrule
    \texttt{ca-init} & Generate the org.\ root CA key and self-signed certificate \\
    \texttt{provision} & Extract the Chip ID and issue the org.\ certificate chain \\
    \texttt{verify} & Match the Chip IDs from the AR/vendor certificates and the org.\ chain (optionally check a CRL) \\
    \texttt{revoke} & Add a certificate serial to the revocation database \\
    \texttt{crl-issue} & Issue an X.509 CRL from the revocation database \\
    \bottomrule
  \end{tabular}
\end{table}
As a CLI, TEE Anchor integrates non-invasively into existing RA verification sequences, for example via subprocesses.

The implementation of TEE Anchor is publicly available at \url{https://github.com/acompany-develop/tee-anchor}.

\section{Evaluation}\label{sec:eval}
Among the commands described in Section~\ref{sec:impl}, we evaluated the operations on SGX Quotes on an Azure SGX instance (machine~B), and all the others on a GCP TDX instance (machine~A). Since TEE Anchor targets fleets managed by diverse organizations, we deliberately avoided locking the experiments into a single vendor. As Verifier tooling, we used an RA script implemented with Intel's official RA library (DCAP) for TDX~\cite{humane-rafw-tdx}, \texttt{snpguest}~\cite{snpguest} for SEV-SNP, and \texttt{evcli}~\cite{evcli} for CCA. For SGX, we used a program implemented against Microsoft Azure Attestation (MAA)~\cite{humane-rafw-maa}, as a test of the case where conventional RA verification is also delegated to a remote Verifier. Since no CCA production hardware was available at the time of writing, the CCA experiments use the CPAK and \texttt{cca-platform-instance-id} provided by a QEMU full-emulation environment~\cite{cca-qemu}. The design remains applicable once such hardware appears. Machine specifications, software versions, and the detailed assignment of experiments to the two machines are given in Appendix~\ref{app:setup}.

Figure~\ref{fig:bench} shows the results. Each measurement was repeated at least 200 times, or more when the benchmarking tool \texttt{hyperfine} deemed it necessary, and we report the mean and standard deviation of the execution times. The additional overhead introduced by TEE Anchor in the attestation verification phase was about 5.2\,ms on TDX (36\% of conventional RA) and about 5.5\,ms on SGX with MAA as the Verifier (4.5\%). In both cases, the overhead was stable at roughly 5\,ms. In the SGX experiment, the relative overhead is smaller because the measurement includes remote communication with MAA.   

We also ran \texttt{verify} under the same conditions in the mode that takes a CRL and performs revocation checking. The overhead was within 1\,ms, or in some cases the execution time was even shorter than without revocation checking, indicating essentially no performance impact. The SEV-SNP and CCA verifications cannot be compared directly in terms of intrinsic overhead because the baselines are tools written in different implementation languages (\texttt{snpguest}, \texttt{evcli}), but both remained below 10\,ms.

Provisioning, a one-time operation at registration, took 5.7--7.1\,ms across all TEE types. CA key generation and revocation processing (\texttt{revoke} + \texttt{crl-issue}) also completed in under 10\,ms. In practice, RA additionally involves AR issuance and transfer to the remote party, so end-to-end times are much larger than under our conditions, as the MAA-based SGX results indicate. TEE Anchor therefore integrates into RA with minor overhead.

\section{Discussion}
\subsection{Comparison with Related Work}
Technologies and research that raise trustworthiness through affiliation assurance existed even before realistic physical attacks such as TEE.fail and Battering RAM emerged, and unsurprisingly more have followed since. We survey such prior work and compare it with TEE Anchor.

\subsubsection{AMD SEV-SNP VLEK}
The Versioned Loaded Endorsement Key (VLEK)~\cite{snp-abi} in AMD SEV-SNP is an AMD-certified AK that can replace the chip-derived VCEK when signing ARs. It is derived from a seed that AMD's Key Derivation Service manages for each cloud service provider (CSP) registered with AMD. The CSP requests a VLEK for a machine designated by its TCB and \texttt{CHIP\_ID}, and once the VLEK is installed, the firmware uses it in place of the VCEK. A VLEK certificate thus binds the origin of an AR to a specific CSP, which is a form of affiliation assurance effective against physical attacks. However, VLEK is in principle intended for CSPs, and it is difficult to read AMD's official documentation~\cite{snp-abi} as opening it to arbitrary organizations. Even if it were open, VLEK applies only to SEV-SNP. In contrast, TEE Anchor can be deployed by any organization, vendor-independently, across TEEs.

\subsubsection{Intel POE}
The prior proposal closest in spirit to TEE Anchor is Intel's Platform Ownership Endorsements (POE)~\cite{poe}, a mechanism Intel proposes as a physical-attack countermeasure that lets a remote verifier confirm who physically owns a platform. The owner collects from each machine the PIID (Platform Instance ID), a unique machine identifier that appears, like the PPID, in the PCK Certificate and in the Platform Manifest used at machine registration. The owner then issues an owner-signed certificate, the POE, that contains this identifier. At verification time, matching the PIID in the PCK Certificate, for instance, against the one in the POE establishes that the Attester belongs to a trustworthy owner. Intel ships POE as part of DCAP~\cite{poe-github}, and unlike VLEK it is expected to be usable by arbitrary organizations.  

POE thus shares its basic structure with our work, namely an owner-signed endorsement over a machine-unique identifier that is matched against vendor evidence. However, Intel officially states that POE is premised on integration into DCAP, which is dedicated to SGX/TDX, and that POE centers on Intel-specific identifiers~\cite{poe}. The draft POE profile for CoRIM likewise indicates SGX/TDX only~\cite{poe-draft}. POE therefore shares VLEK's lack of a deployment path to other vendors' TEEs, whereas TEE Anchor uses only the common Chip ID abstraction and standard X.509 to handle SGX, TDX, SEV-SNP, and CCA uniformly.

\subsubsection{DCEA and Dstack}
Even before TEE.fail was disclosed, the authors who advocated CSP-managed allowlists of acceptable PPIDs as a physical-attack countermeasure for current TEEs~\cite{narrowing} proposed Data Center Execution Assurance (DCEA)~\cite{dcea}. In DCEA, the TEE's AR authenticates the AK of a vTPM deployed on the host outside the TEE. Some variants additionally use the Quote of the machine's dTPM. The CSP then issues a certificate for the Endorsement Key (EK), the key superior to the AK in the TPM hierarchy, which proves that the TEE bound to that vTPM belongs to the CSP. Dstack~\cite{dstack} discusses a similar approach combining PPID checks with TPM-based binding.  

Unlike TEE Anchor, these approaches depend on a TPM for their trust guarantees. This adds TPM-specific attack vectors entirely separate from the TEE, such as TPM-Fail~\cite{tpmfail}, which requires no physical access yet strikes at the core of TPM attestation. It also incurs the non-negligible overhead of TPM access, as the DCEA paper reports about 550\,ms for a dTPM Quote and 300\,ms for a vTPM Quote~\cite{dcea}, whereas every TEE Anchor operation stays below 10\,ms. TEE Anchor's TEE-and-PKI-only design is thus advantageous in both security and performance.

\subsubsection{Proof of Cloud Alliance}
Proof of Cloud Alliance~\cite{poc} registers verified correspondences between Chip IDs and physical locations in a signed, append-only registry jointly managed by multiple independent organizations. Like our work, it uses Chip IDs for affiliation assurance. However, registering a machine requires costly measures such as physical witnessing by members, zk-TLS proofs, tamper-evident RFID beacons, and the aforementioned DCEA. Additions to the registry further require a majority vote, and both the Chip ID allowlist and the revocation list must be fetched from the centrally managed database~\cite{trustserver}.  

With TEE Anchor, in contrast, each organization independently decides which machines it trusts by issuing certificates from its own CA, without any shared registry, multi-party consensus, or heavyweight verification procedures. Whereas the Alliance aims at a shared registry among mutually distrusting participants, TEE Anchor lets any single organization introduce affiliation assurance independently and cheaply, differentiating the two in applicable domains and operational cost.

\subsubsection{SPIFFE/SPIRE}
SPIFFE/SPIRE issues attestation-based identities (SVIDs) to workloads, and Pontes et al.\ proposed a SPIRE plugin for SEV-SNP CVMs~\cite{spire}. In their scheme, a SPIRE Server operated by the organization performs RA against the TEE, issues an SVID, and hands it to downstream Verifiers. Though not proposed as a physical-attack countermeasure, it proves organizational affiliation and could thus be applied, like TEE Anchor, as an indirect defense. In this scheme, however, the downstream Verifier merely trusts the SVID without performing conventional RA itself, a model that, like MAA, entrusts even the execution of RA to a central Verifier. TEE Anchor instead has the Verifier (RP) perform conventional RA itself and layer Chip ID matching on top, which minimizes the trust placed in the organization. The two are not substitutes but complementary designs that differ in trust model, with SPIRE concentrating trust in a central server and TEE Anchor leaving it with each Verifier.

\subsection{Mitigating Attacks on Attested TLS}\label{sec:attls}
Two main attacks have been discussed against Attested TLS, a class of constructions that bind RA evidence to the TLS session between the RP and the Attester (see Appendix~\ref{app:attls} for background).The first is the diversion attack~\cite{diversion}. When two distinct TEE instances with identical measurements and metadata run on entirely different machines, a connection can be silently steered to the unintended instance. The second is a relay attack on the session binding~\cite{relay}, which is distinct from the proxy/relay attacks of Section~\ref{sec:proxyrelay}. The attacker first establishes an Attested TLS session with the Attester. The attacker then reuses the session-specific information bound into the Report Data, a user-chosen field in the AR, in its own Attested TLS session with the RP, and can thereby read both channels unnoticed.

We argue that TEE Anchor can be used to mitigate both, taking the following positions in line with the standard TEE threat model: (1)~a TEE instance whose measurements and related metadata have been deemed trustworthy via RA is uniformly regarded as trustworthy; (2)~however, it has no inherent resistance to physical attacks; (3)~we exclude leakage of an Attester TEE instance's private keys by means other than physical attacks, because workload bugs are outside the TEE's scope, and TCB exploits are either detectable via RA or, if unknown zero-days, compromise the TEE itself and render the discussion moot; and (4)~the Attester's server side enforces that no weak algorithms are used in the handshake, which is verifiable via RA as part of the measurements.

Under these assumptions, a diversion attack is a threat only when it steers the connection to a machine on which physical attacks are feasible, and TEE Anchor rules this out by guaranteeing the Attester machine's affiliation. As for relay attacks, robust Attested TLS constructions bind session-specific data such as the server certificate into the Report Data, unlike weak ones that bind only a nonce. Against them, the attack can succeed only if the server private key paired with the bound certificate, or the AK private key, leaks from the TEE or the TCB for use in forged signatures. Under assumptions (2) and (3), such leakage requires a physical attack. Consequently, proving with TEE Anchor that the machine belongs to a trustworthy organization where physical attacks are infeasible also serves as a countermeasure against relay attacks. Relay attacks are also expected to be preventable by the post-handshake style of Attested TLS~\cite{post-hs}, but TEE Anchor mitigates physical and relay attacks simultaneously.

\section{Conclusion}
This paper proposed TEE Anchor, a lightweight mechanism guaranteeing that a TEE machine physically resides under trustworthy management. Physical attacks such as TEE.fail and Battering RAM defeat current TEEs, and proving affiliation mitigates them indirectly. Conventional RA, however, proves \emph{what code is running} but not \emph{where}. Treating affiliation assurance as a new security requirement, TEE Anchor embeds the CPU-unique Chip ID into X.509 extensions signed by an organizational CA, defeating proxy/relay attacks without additional roots of trust such as (v)TPMs. It differs from existing approaches in covering SGX, TDX, SEV-SNP, and CCA through one abstraction, requiring no trust in central verification services, and letting each organization issue and revoke certificates independently. Our prototype evaluation confirmed additional verification overhead of a few milliseconds. We further discussed how the same guarantee can be used to mitigate diversion and relay attacks on Attested TLS.  

The affiliation claim's authenticity relies on trust in the organization's physical security, while all other attestation is verified directly against the TEE vendor's root of trust. Future work includes operational deployment and resettable identifiers for supply-chain resilience.

%% NOTE: acmart suppresses the acks environment when the "anonymous" option
%% is active, so this appears only in the camera-ready version. The AI-usage
%% disclosure below (per the ACM Policy on Authorship, as required by the
%% SysTEX CFP) is therefore duplicated in Appendix A so that it is visible
%% to reviewers in the anonymous submission as well.
\begin{acks}
Generative AI tools (large language models) were used in this work to assist with the implementation of the prototype and with the translation and editing of this manuscript, which was originally drafted in Japanese. All technical designs, experiments, and claims were conceived, conducted, and verified by the authors, who take full responsibility for the content of this paper.
\end{acks}

%% \sloppy + emergencystretch inside the bibliography only: lets TeX accept
%% slightly loose lines instead of letting long URLs protrude into the margin.
\bibliographystyle{ACM-Reference-Format}
{\sloppy\emergencystretch=1.5em
\bibliography{refs}}

@techreport{ccc-wp,
  author      = {Patrick Bues},
  title       = {Unlocking the Future of Data Security: Confidential Computing as a Strategic Imperative},
  institution = {Confidential Computing Consortium},
  type        = {White Paper},
  year        = {2025},
  url         = {https://confidentialcomputing.io/wp-content/uploads/sites/10/2025/11/US53866125.pdf}
}

@misc{snp-abi,
  author       = {{AMD}},
  title        = {{SEV} Secure Nested Paging Firmware {ABI} Specification},
  howpublished = {Revision 1.58},
  year         = {2025},
  url          = {https://www.amd.com/content/dam/amd/en/documents/developer/56860.pdf}
}

@misc{poe,
  author       = {{Intel}},
  title        = {Platform Ownership Endorsements for Confidential Computing},
  year         = {2026},
  url          = {https://www.intel.com/content/www/us/en/developer/articles/technical/software-security-guidance/technical-documentation/platform-ownership-endorsements.html}
}

@inproceedings{teefail,
  author    = {Jesse Chuang and Adam Seto and Nicolas Berrios and Stephan van Schaik and Christina Garman and Daniel Genkin},
  title     = {{TEE.fail}: Breaking Trusted Execution Environments via {DDR5} Memory Bus Interposition},
  booktitle = {Proceedings of the 47th IEEE Symposium on Security and Privacy (S\&P)},
  year      = {2026},
  url       = {https://tee.fail}
}

@inproceedings{batteringram,
  author    = {Jesse {De Meulemeester} and David Oswald and Ingrid Verbauwhede and Jo {Van Bulck}},
  title     = {Battering {RAM}: Low-Cost Interposer Attacks on Confidential Computing via Dynamic Memory Aliasing},
  booktitle = {Proceedings of the 47th IEEE Symposium on Security and Privacy (S\&P)},
  month     = may,
  year      = {2026},
  url       = {https://batteringram.eu/}
}

@inproceedings{narrowing,
  author    = {Filip Rezabek and others},
  title     = {Narrowing the Gap between {TEEs} Threat Model and Deployment Strategies},
  booktitle = {Proceedings of the 8th Workshop on System Software for Trusted Execution (SysTEX '25)},
  year      = {2025},
  url       = {https://systex-workshop.github.io/2025/papers/systex25-final1.pdf}
}

@misc{dcea,
  author        = {Filip Rezabek and Moe Mahhouk and Andrew Miller and Quintus Kilbourn and Georg Carle and Jonathan Passerat-Palmbach},
  title         = {Proof of Cloud: Data Center Execution Assurance for Confidential {VMs}},
  howpublished  = {arXiv preprint arXiv:2510.12469},
  year          = {2025}
}

@inproceedings{tpmfail,
  author    = {Daniel Moghimi and Berk Sunar and Thomas Eisenbarth and Nadia Heninger},
  title     = {{TPM-FAIL}: {TPM} meets Timing and Lattice Attacks},
  booktitle = {Proceedings of the 29th USENIX Security Symposium (USENIX Security 20)},
  pages     = {2057--2073},
  year      = {2020}
}

@inproceedings{spire,
  author    = {Davi Pontes and Fernando Silva and Eduardo Falc{\~a}o and Andrey Brito},
  title     = {Attesting {AMD} {SEV-SNP} Virtual Machines with {SPIRE}},
  booktitle = {Proceedings of the 12th Latin-American Symposium on Dependable and Secure Computing (LADC '23)},
  pages     = {1--10},
  year      = {2023},
  url       = {https://doi.org/10.1145/3615366.3615419}
}

@inproceedings{diversion,
  author    = {Muhammad Usama Sardar and Mariam Moustafa and Tuomas Aura},
  title     = {Identity Crisis in Confidential Computing: Formal Analysis of Attested {TLS}},
  booktitle = {Proceedings of the ACM Asia Conference on Computer and Communications Security (ASIA CCS '26)},
  publisher = {Association for Computing Machinery},
  address   = {New York, NY, USA},
  pages     = {547--560},
  year      = {2026}
}

@misc{relay,
  author       = {Muhammad Usama Sardar and Vyacheslav Dubeyko and Jean-Marie Jacquet},
  title        = {Intra-handshake.fail ({CVE-2026-33697}): High-severity {CVE} in Attested {TLS}},
  year         = {2026}
}

@misc{rats,
  author       = {{IETF}},
  title        = {Remote ATtestation ProcedureS (rats)},
  note         = {Accessed: 2026-06-17},
  url          = {https://datatracker.ietf.org/group/rats/about/}
}

@misc{poc,
  author       = {{Proof of Cloud Alliance}},
  title        = {Proof of Cloud},
  note         = {Accessed: 2026-06-25},
  url          = {https://proofofcloud.org/}
}

@misc{poe-draft,
  author       = {{IETF}},
  title        = {A {CoRIM} Profile for {Intel} Platform Ownership Endorsements ({POE})},
  note         = {Internet-Draft draft-bzb-rats-intel-poe-endorsements. Accessed: 2026-08-31},
  year         = {2026},
  url          = {https://datatracker.ietf.org/doc/draft-bzb-rats-intel-poe-endorsements/}
}

@misc{post-hs,
  author       = {{IETF}},
  title        = {Remote Attestation with Exported Authenticators},
  note         = {Internet-Draft draft-fossati-seat-expat. Accessed: 2026-09-04},
  year         = {2026},
  url          = {https://datatracker.ietf.org/doc/draft-fossati-seat-expat/}
}

@misc{humane-rafw-tdx,
  author       = {{Acompany Co., Ltd.}},
  title        = {{Humane-RAFW-TDX}/relying-party/rp\_client.py},
  note         = {Accessed: 2026-06-23},
  url          = {https://github.com/acompany-develop/Humane-RAFW-TDX/blob/main/relying-party/rp_client.py}
}

@misc{humane-rafw-maa,
  author       = {{Acompany Co., Ltd.}},
  title        = {{Humane-RAFW-MAA}/Client\_App/client\_app.cpp},
  note         = {Accessed: 2026-06-23},
  url          = {https://github.com/acompany-develop/Humane-RAFW-MAA/blob/main/Client_App/client_app.cpp}
}

@misc{snpguest,
  author       = {{virtee}},
  title        = {snpguest},
  note         = {Accessed: 2026-06-23},
  url          = {https://github.com/virtee/snpguest}
}

@misc{evcli,
  author       = {{veraison}},
  title        = {evcli --- Attestation Evidence Manipulation Tool},
  note         = {Accessed: 2026-06-23},
  url          = {https://github.com/veraison/evcli}
}

@misc{cca-qemu,
  author       = {{Linaro}},
  title        = {Building a {CCA} Stack for {QEMU} (cca/v10)},
  note         = {Accessed: 2026-06-23},
  url          = {https://gitlab.com/Linaro/cca-public/build-instructions}
}

@misc{dstack,
  author       = {{Dstack-TEE}},
  title        = {{TEE}: {TPM}-Based Approach to Code and Location Verification \#626},
  note         = {GitHub Discussion. Accessed: 2026-06-25},
  url          = {https://github.com/Dstack-TEE/dstack/discussions/626}
}

@misc{trustserver,
  author       = {{Proof of Cloud Alliance}},
  title        = {Confidential {VM} Quote Processing Server},
  note         = {Accessed: 2026-06-25},
  url          = {https://github.com/proofofcloud/trust-server}
}

@misc{poe-github,
  author       = {{Intel}},
  title        = {intel/confidential-computing.tee.dcap.poe},
  note         = {GitHub repository. Accessed: 2026-08-31},
  url          = {https://github.com/intel/confidential-computing.tee.dcap.poe}
}

\appendix
\section{Background on Attested TLS}\label{app:attls}
Attested TLS gives the RP confidence that its encrypted TLS session with the Attester terminates at a TEE the RP has deemed trustworthy. To achieve this, it performs RA and binds the resulting evidence to that TLS session. Various protocols exist. A common one binds the session by embedding session-specific information, such as the TLS certificate, into the AR's Report Data, a field whose contents the user can freely specify.

\section{Experimental Setup Details}\label{app:setup}
Machine~A is a TDX-enabled c3-standard-4 on GCP (4~vCPUs, 16\,GB RAM) running Ubuntu 24.04 LTS with Linux-SGX v2.29, DCAP v1.26, snpguest v0.10.0, and evcli v2.2.0+dirty (ccatoken v1.4.0, go-cose v1.3.0). Machine~B is an Azure Standard DC4s~v3 (4~vCPUs, 32\,GB RAM); its Ubuntu, Linux-SGX, and DCAP versions are identical to machine~A.

\texttt{ca-init}, \texttt{revoke}, and \texttt{crl-issue} are independent of the TEE type, so we ran them only on machine~A. For \texttt{provision} and \texttt{verify}, we prepared the ARs and TEE vendor certificates for each of the four TEEs and ran the SGX experiments on machine~B and the others on machine~A. Although provisioning should normally be performed on the TEE machine itself, we moved the SEV-SNP and CCA ARs and vendor certificates onto machine~A to minimize performance variation across machines. For \texttt{verify} in the presence of a revocation list, since revocation processing itself does not depend on the TEE type, we ran the TDX and SGX experiments on machines~A and~B respectively, solely to absorb per-machine variation.

\section{Open Science}
The primary research artifact of this paper is \texttt{tee-anchor}, the standalone CLI tool described in Section~\ref{sec:impl}. It implements the provisioning, verification, and CA lifecycle operations of TEE Anchor, where the CA lifecycle operations comprise revocation and CRL issuance, for Intel Scalable-SGX/TDX, AMD SEV-SNP, and Arm CCA. The tool, together with the benchmark scripts and the raw measurement data underlying Figure~\ref{fig:bench}, is available as open source at \url{https://github.com/acompany-develop/tee-anchor}.

The evaluation additionally relies on third-party open-source tools, namely \texttt{snpguest} and \texttt{evcli}, and on publicly documented cloud instance types, namely GCP c3-standard-4 with TDX and Azure DC4s~v3 with SGX. The experiments can therefore be reproduced without access to proprietary infrastructure. The CCA experiments use the publicly available QEMU-based CCA emulation stack.

\end{document}